# Twin GEM-TPC Prototype (HGB4) Beam Test at GSI – a Development for the Super-FRS at FAIR

F. García[*1], R. Turpeinen[1], R. Lauhakangas[1], E. Tuominen[1], J. Heino[1], J. Äystö[1], T. Grahn[2], S. Rinta-Antilla[2], A. Jokinen[2], R. Janik[3], P. Strmen[3], M. Pikna[3], B. Sitar[3], B. Voss[4], J. Kunkel[4], V. Kleipa[4], A. Gromliuk[4], H. Risch[4], I. Kaufeld[4], C. Caesar[4], C. Simon[4], M. kìs[4], A. Prochazka[4], C. Nociforo[4], S. Pietri[4], H. Simon[4], C. J. Schmidt[4], J. Hoffmann[4], I. Rusanov[4], N. Kurz[4], P. Skott[4], S. Minami[4], M. Winkler[4]

[1]Helsinki Institute of Physics, University of Helsinki, 00014 Helsinki, Finland
[2]University of Jyväskylä, Department of Physics, 40014 Jyväskylä, Finland
[3]FMFI Bratislava, Comenius University, Bratislava, Slovakia
[4]GSI Helmholtzzentrum für Schwerionenforschung, Darmstadt 64291, Germany

## INTRODUCTION

The GEM-TPC detector will be part of the standard Super-FRS detection system, as tracker detectors at several focal stations along the separator and its three branches.

## GEM-TPC DETECTOR DEVELOPMENT

In order to satisfy the requirements of the Super-FRS a GEM-TPC working group was created. From 2009 different designs were proposed and after two generations of GEM-TPCs [1] were built and tested, one of the main requirements, which was not yet completely tested was a close to 100% tracking efficiency at high rates, therefore the twin configuration was introduced. The main idea behind this new design is to place two GEM-TPCs one close to the other one and flipped in the middle horizontal plane, in such a way that the electric field of the field cages will be in opposite directions. Results from simulations indicate that this configuration can potentially achieve a close to 100% tracking efficiency at up to 1 MHz rate.

## BEAM TEST AT GSI

The twin prototype called HGB4, shown in Fig. 1 was fully designed and assembled at the GSI Detector Laboratory in 2014.

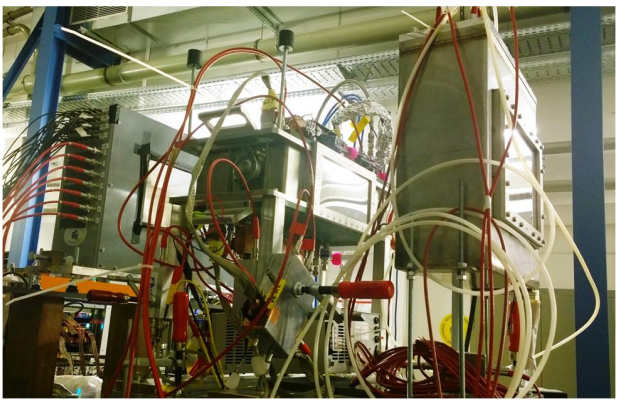

Fig. 1. The HGB4 (midlle) at Cave-C during in beam test.

After a successful commissioning, the chamber was moved to Cave-C for beam tests. This detector was tested by using Ca, Bi and U beams from SIS18 of about 300 MeV/u. This campaign was used to carry out measurements of the Control Sum (c.s.) for different electric fields.

$$c.s. = T_{up} + T_{down} - 2T_{ref}$$

Where: $T_{up} + T_{down}$ is the sum drift times for both field cages and $T_{ref}$ is the reference time from the plastic scintillator. In order to measure the drift time the signal from the bottom of the third GEM was picked up. The c.s. was measured by using a multihit TDC (Caen V1290) for different fields starting from 150 V/cm up to 320 V/cm.
We measured the Control Sum for different fields starting from 150 V/cm up to 320 V/cm.

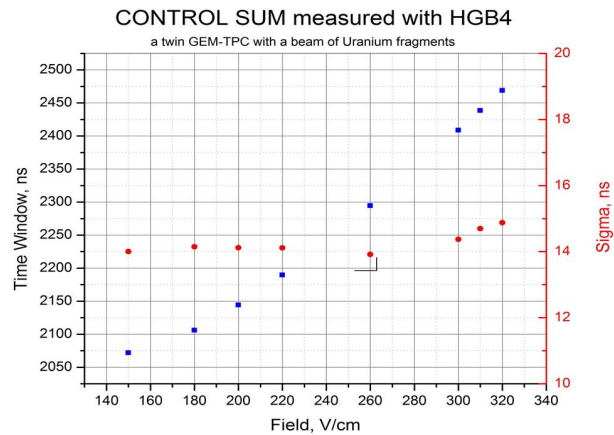

Fig. 2. The HGB4 Control Sum (blue dots) and its sigma (red dots).

From the Fig. 2 one can see that the sigma of the c.s. distribution slightly changes with different fields, indicating the possibility to operate at lower field without degradation, however more test are planned in the future to confirm these results.